# Results from a Prototype Proton-CT Head Scanner

R.P. Johnson[a]\*, V.A. Bashkirov[b], G. Coutrakon[c], V. Giacometti[d], P. Karbasi[e],
N.T. Karonis[c], C.E. Ordoñez[c], M. Pankuch[f], H.F.-W. Sadrozinski[a], K.E. Schubert[e],
R.W. Schulte[b]

[a]*U.C. Santa Cruz Physics Department, 1156 High St., Santa Cruz, CA 95065, USA*
[b]*Division of Radiation Research, Loma Linda University, Loma Linda, CA 92354, USA*
[c]*Departments of Physics and Computer Science, Northern Illinois University, DeKalb, IL 60115, USA*
[d]*Centre for Medical Radiation Physics, University of Wollongong, Wollongong, NSW, Australia*
[e]*School of Engineering and Computer Science, Baylor University, Waco, TX 76798, USA*
[f]*Northwestern Medicine Chicago Proton Center, 4455 Weaver Parkway, Warrenville, IL 60555, USA*

**Abstract**

We are exploring low-dose proton radiography and computed tomography (pCT) as techniques to improve the accuracy of proton treatment planning and to provide artifact-free images for verification and adaptive therapy at the time of treatment. Here we report on comprehensive beam test results with our prototype pCT head scanner. The detector system and data acquisition attain a sustained rate of more than a million protons individually measured per second, allowing a full CT scan to be completed in six minutes or less of beam time. In order to assess the performance of the scanner for proton radiography as well as computed tomography, we have performed numerous scans of phantoms at the Northwestern Medicine Chicago Proton Center including a custom phantom designed to assess the spatial resolution, a phantom to assess the measurement of relative stopping power, and a dosimetry phantom. Some images, performance, and dosimetry results from those phantom scans are presented together with a description of the instrument, the data acquisition system, and the calibration methods.

\* Corresponding author. Tel.: +1-831-459-2125; fax: +1-831-459-5777.
 E-mail address: rjohnson@ucsc.edu

## 1. Introduction

Proton computed tomography (pCT) is a technology under development that holds promise to improve hadron therapy treatment planning by making direct measurements of proton stopping power that have up to now been estimated from X-ray absorption measurements. The concept of pCT was proposed as early as 1963 (Cormack, 1963), but no system for pCT yet exists for use on patients. Modern technology has allowed us to realize a prototype pCT scanner system that is compact, reliable, of reasonable cost, and capable of acquiring in six minutes or less the data for a high quality, accurate, low-artifact 3-dimensional CT image of proton relative stopping power (RSP) in a subject the size of a human head. In this paper the scanner hardware is briefly introduced and some of our recent results from CT scans of phantoms are presented.

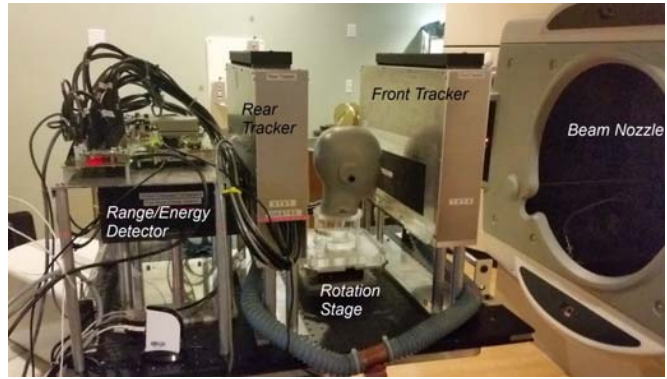

Figure 1. Photograph of the pCT scanner located in a fixed-beam treatment room of the Northwestern Medicine Chicago Proton Center, with a pediatric head phantom (model 715-HN from CIRS Inc., Norfolk, VA) mounted on the rotation stage.

## 2. The Scanner

The scanner hardware has been thoroughly described elsewhere (Johnson, 2016), so the description included here is brief. See Fig. 1 for a photograph. The scanner measures protons one by one at a sustained rate of greater than a million per second. Each measurement consists of a position and direction for the proton entering the phantom ("Front Tracker"), another position and direction for the proton as it exits ("Rear Tracker"), and the residual range of the exiting proton in polystyrene scintillator ("Range/Energy Detector"). The tracking measurements are used to find the "most likely path (MLP)" of the proton through the phantom (Schulte, 2008), while from the residual range together with the known beam energy, the "water equivalent path length (WEPL)" of the proton through the phantom is inferred. The set of WEPL values and associated MLPs through the object over a 360° angular scan is processed by an iterative parallelizable reconstruction algorithm that runs on modern GP-GPU hardware. A detailed Monte Carlo simulation model also exists that accurately reproduces the scanner's performance (Giacometti, 2017), which allows us to perform studies to understand the ways in which all aspects of the design impact the instrument's performance.

At the Northwestern Medicine Chicago Proton Center, several weekend runs in a 200 MeV proton beam[†] have been completed with scans made of several phantoms. Each scan normally lasted for six minutes, with six full rotations made of the phantom on a computer controlled stage located between the two tracking detectors. The event rate of around one MHz results in a file of over 300 million events in over 20 Gbytes of raw data. A software program that is separate from the data acquisition unpacks the raw data and reconstructs the tracks and calibrated

---

[†]The 230 MeV beam from the cyclotron is degraded to 200 MeV with a one-sigma spread of about 1 MeV. About 3 MeV in addition is lost from passing through windows and silicon detectors.

WEPL, which are output to a file of "proton histories" to be used by image reconstruction. That processing has been accomplished in 3.5 minutes on a fairly modest workstation running 12 parallel threads, so it can be done essentially in parallel with the data acquisition.

Each of the two tracking detectors has four layers of silicon strip detectors, two measuring the horizontal coordinates and two measuring the vertical coordinates. The detectors operate with nearly zero noise and with greater than 99% efficiency for protons passing through active silicon, which has an aperture of about 9 cm by 36 cm. The range/energy detector is implemented as five "stages" of scintillator, each 5.1 cm thick and read out by a photomultiplier tube (PMT). The signals are digitized to give an accurate measurement of the light yield, which essentially provides a measurement of how far the proton penetrates into the stage in which it stops. A detailed presentation of this concept has been published elsewhere (Bashkirov, 2016). The data acquisition is a fully custom implementation based on integrated circuits designed specifically for this application to read out the silicon strips (Johnson, 2013) and a hierarchy of 15 field programmable gate arrays (FPGA) programmed in Verilog. The final FPGA, a Xilinx Vertex-6 device, acts as the event builder, which sends the data over a single 100 Mbyte/s Ethernet link to the data acquisition computer. In the present implementation the Ethernet link limits the event rate to a maximum of about 1.2 to 1.3 MHz. At a 1.3 MHz event rate the trigger rate is found to be 1.6 MHz, for a dead time fraction of 25%, so normally the scanner is operated at an event rate closer to 1 MHz, for which the dead time fraction is less than 10%.

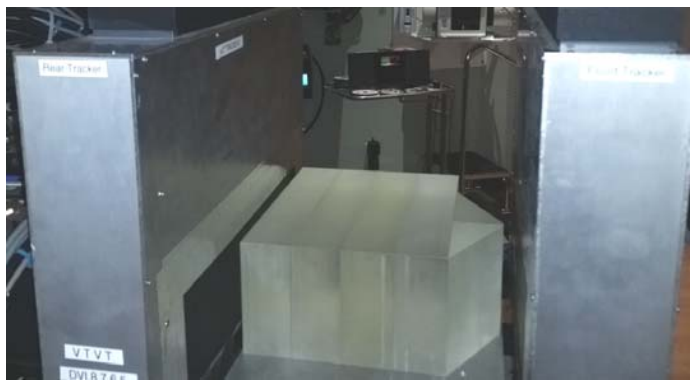

Figure 2. The new WEPL calibration phantom installed between the tracking detectors. The four rectangular blocks are removable, to provide the full WEPL range needed for calibration.

## 3. WEPL Calibration

Whereas the silicon-strip tracking detectors require essentially no calibration beyond initial alignment and setting of thresholds, successful use of the five-stage energy/range detector relies on a fairly complex calibration procedure that is executed at the beginning of each run period. The procedure as originally implemented is described in detail elsewhere (Bashkirov, 2016). It is time consuming, as it relies on personnel going in and out of the beam area to place four plastic blocks for successive runs. However, each of the runs requires less than ten seconds, so it would be straightforward to implement a fast procedure by robotic placement of the blocks.

The first images published (Johnson, 2016) suffered from obvious ring-like artifacts that resulted from inadequate calibration. Since then the calibration has been greatly improved by implementation of a new calibration phantom and procedure plus implementation of real-time PMT gain tracking during data processing, accomplished by analysis of protons that miss the phantom and therefore enter the energy/range detector with full energy. The new calibration phantom can be seen in Fig. 2. In practice six runs are taken, one with no phantom, used to calibrate and flatten the response across the aperture, one with just the wedges, and then four more with successively more blocks in place. The phantom does not cover the entire detector aperture, such that in every run there are many full-energy protons entering the energy/range detector for gain normalization. From the last five runs a set of five calibration curves is derived, as illustrated in Fig. 3. For the figure, each curve was derived from data of only those protons that appeared to stop in the given stage. Given the curve for the stage in which the proton stops, the data processing program uses it to translate from the measured stage energy to the WEPL value. As can be seen in the images presented below, the new calibration provides artifact-free images.

## 4. Image Reconstruction and Scanner Performance

Because of the complexity of taking into account the nonlinear MLP, the image reconstruction is done by an iterative projection method, using an algorithm that is suitable for parallelization, rather than by a filtered back-projection (FBP). Details and performance results can be found in the CAARI presentation by Caesar Ordoñez (Ordoñez, 2017). It is a challenge to complete the image reconstruction in a time interval comparable to the six-minute data acquisition, but Ordoñez demonstrated that it can be accomplished in only 4.5 minutes using ten compute nodes, each with twelve cores and two GPUs. Similarly, using different code Schultze demonstrated (Schultze, 2015) that a similar image could be reconstructed in 6.5 min on a single Xeon compute node using one NVIDIA K40 GPU.

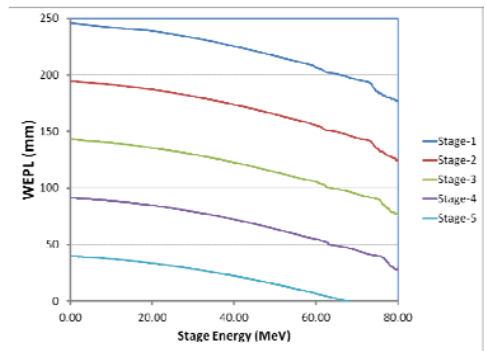

Figure 3. WEPL calibration curves for 200 MeV protons.

Figure 4 shows a reconstructed image of a CTP404 Geometry and Sensitometry module called the "sensitometry" phantom, a plastic cylinder with an assortment of cylindrical inserts (Catphan$^{TM}$ 504 phantom by The Phantom Laboratory, Salem, NY). It can be compared with an earlier published image (Johnson, 2016) that suffered from obvious ring artifacts. Here the calibration is of sufficient quality that any remaining artifacts are below the level of visibility. Each of the inserts in this phantom is made of a different material, with RSP measured with the PeakFinder$^{TM}$ instrument (PTW GmbH, Freiburg, Germany), allowing the accuracy of the measurement of proton RSP to be assessed. In fact, from the reconstruction shown in Fig. 4, all of the RSP values are reconstructed to better than 1% accuracy (Ordoñez, 2017). The edges of the inserts appear fuzzy due to the spatial resolution, which is limited by multiple scattering of the protons in the phantom. The spatial resolution of the system was thoroughly evaluated by scanning a custom phantom with cubic inserts. The details are published elsewhere (Plautz, 2016). Figure 5 shows reconstructed images of a realistic pediatric head phantom.

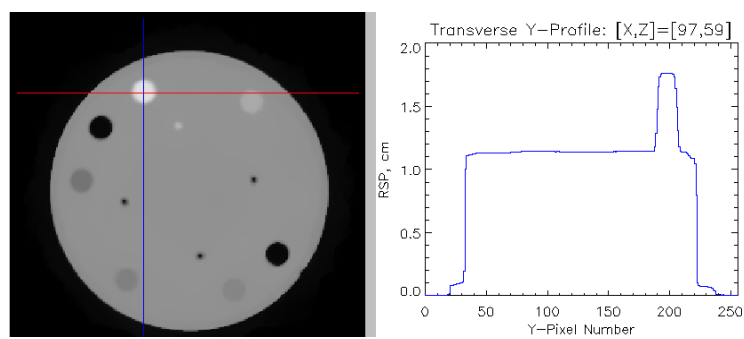

Figure 4. Reconstructed slice in the sensitometry phantom (left) and a profile (right) taken along the vertical blue line in the slice.

## 5. Radiological Dose

Proton CT has long been anticipated to be a low-dose imaging modality, so it is important to verify that expectation. We measured the dose during a typical 6-minute CT scan by using a Farmer ionization chamber (Model N30013 by PTW GmbH, Freiburg, Germany) inserted into the center of a Catphan$^{TM}$ CTP554 acrylic dose phantom, 16 cm in diameter (Giacometti, 2016). At the same time the scanner tracking system measured the protons entering and exiting the phantom, from which we inferred the proton flux in the central region of the phantom. The measured dose was 1.39 mGy (water equivalent), for which the estimated flux was 1.2 million protons per cm$^2$. Estimating the ionization in water that would result from that flux accounts for about 80% of the dose measured by the ionization chamber, which is reasonable considering the complexities of extrapolating the tracking measurements to the phantom center and properly accounting for inefficiencies. A typical dose delivered by an X-ray CT scan of the head is 30–50 mGy (Mettler, 2008), so the scans that produced Fig. 4 and other similar images (Ordoñez, 2017) correspond to doses considerably lower than that of a typical X-ray CT dose. We have not yet ascertained the optimal number of proton histories needed for effective treatment planning. The 6-minute duration was chosen to achieve a scan with very low noise content (<1%). In practice the dose can most likely be reduced significantly to 1 mGy or less for a full pCT scan and around 0.01 mGy for a single radiograph. The corresponding effective dose will most likely be in the micro-Sv range, but confirmation of that will require a careful radiobiological evaluation.

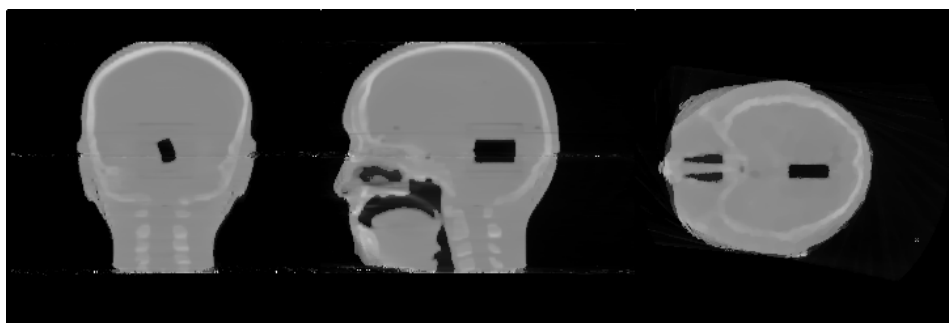

Figure 5. Three slices from a pCT reconstruction of a customized pediatric head phantom (Model 715HN by CIRS Inc., Norfolk, VA). Two six-minute scans, one above the other, were combined to make the images. The rectangular air-filled hole (dark area) in the head is the cavity for a dosimetry-film holder.

## 6. Conclusions

A pCT scanner with an aperture large enough to scan a human head has been built and successfully tested on a variety of phantoms. The system is able to make a complete CT scan in six minutes or less, or a single radiograph in a few seconds. The WEPL calibration is now well understood, allowing reconstruction of phantoms with a negligible level of artifacts. The RSP reconstruction accuracy is generally better than 1%, and the spatial resolution is good, although of course not as sharp as in modern X-ray CT because of multiple scattering by the protons. The radiological dose for a 6-minute CT scan has been measured to be only 1.4 mGy, significantly less than what is typical for a similar X-ray CT scan.

## Acknowledgements

The authors acknowledge the crucial support for this project by the technical staff at Baylor, LLU, NIU, and UCSC, as well as the support of the accelerator operators and other staff at the Northwestern Medicine Chicago Proton Center. They also acknowledge Robert Jones at Inland Technical Service for mechanical engineering and machining support. This work was supported in part by the National Institute of Biomedical Imaging and Bioengineering (NIBIB) and the NSF, award R01EB013118. The content is solely the responsibility of the authors and does not necessarily represent the official views of NIBIB and NIH.

## References

Bashkirov, V.A. et al., 2016, Novel Scintillation Detector Design and Performance for Proton Radiography and Computed Tomography, Medical Physics 43-2, 664–674.

Cormack, A.M., 1963, Representation of a Function by Its Line Integrals, with Some Radiological Applications, Journal of Applied Physics 34, 2722.

Giacometti, V. et al., 2016, Dosimetric Evaluation of Proton CT using a Prototype Proton CT Scanner, proceedings of the IEEE NSS/MIC, Oct. 29, 2016 Strasbourg, France.

Giacometti, V. et al., 2017, Software Platform for Simulation of a Prototype Proton CT Scanner, Medical Physics, in press.

Johnson, R.P. et al., 2013, Tracker Readout ASIC for Proton Computed Tomography Data Acquisition", IEEE Transactions on Nuclear Science 60-5, 3262–3269.

Johnson, R.P. et al., 2016, A Fast Experimental Scanner for Proton CT: Technical Performance and First Experience with Phantom Scans, IEEE Transactions on Nuclear Science 63-1, 52–60.

Mettler F.A. Jr. et al., 2008, Effective Doses in Radiology and Diagnostic Nuclear Medicine: A Catalog. Radiology 248-1, 254–263.

Ordoñez, C.E., 2017, A Real-Time Image Reconstruction System for Particle Treatment Planning Using Proton Computed Tomography (pCT), Proceedings of the 24th Conference on Application of Accelerators in Research and Industry, Fort Worth, TX, October 2016.

Plautz, T.E. et al., 2016, An Evaluation of Spatial Resolution of a Prototype Proton CT Scanner, Medical Physics 43, 6291–6300.

Schulte, R.W., Penfold, S.N., Tafas, J.T., and Schubert, K.E., 2008, A Maximum Likelihood Proton Path Formalism for Application in Proton Computed Tomography, Medical Physics 35, 4849.

Schultze, B. et al., 2015, Reconstructing Highly Accurate Relative Stopping Powers in Proton Computed Tomography, proceedings of the IEEE NSS/MIC, Oct. 31, 2015, San Diego, CA.